\documentclass{tlp}
\usepackage{amsmath}
\usepackage{amssymb}
\usepackage[pdftex]{graphicx}
\DeclareGraphicsExtensions{.jpg, .png, .mps, .bmp, .pdf}
\usepackage{epsfig}
\usepackage{subfigure}
\newtheorem{Definition}{Definition} 
\newtheorem{Theorem}{Theorem} 
\begin{document}
\title{
Query Evaluation and Optimization in the Semantic Web}
\author[Edna Ruckhaus, Eduardo Ruiz, Mar\'{\i}a-Esther Vidal]
{Edna Ruckhaus
\and Eduardo Ruiz
\and Mar\'{\i}a-Esther Vidal\\
Computer Science Department\\
Universidad Sim\'on Bol\'{\i}var\\
\email{\{ruckhaus, eruiz, mvidal\}@ldc.usb.ve}
 }

\maketitle
\begin{abstract}
We address the problem of answering Web ontology queries efficiently.
An ontology is formalized as a {\it Deductive
Ontology Base} (DOB), a deductive database that comprises the
ontology's inference axioms and facts. A cost-based 
query optimization technique for DOB is presented. 
A hybrid cost model is proposed to estimate the cost
and cardinality of basic and inferred facts. 
Cardinality and cost of inferred facts are estimated using
an adaptive sampling technique, while techniques of
traditional relational cost models are used for estimating the cost of basic facts and 
conjunctive ontology queries. Finally, we implement a dynamic-programming optimization
algorithm to identify query evaluation plans that minimize the
number of intermediate inferred facts.
We modeled a subset of the Web ontology language OWL Lite as a DOB,
and performed an experimental study to analyze the predictive capacity of our
cost model and the benefits of the query optimization
technique. Our study has been conducted over synthetic and real-world
OWL ontologies, and shows that the techniques are accurate and improve query performance. To appear in Theory and
Practice of Logic Programming (TPLP)

\end{abstract}
\section{Introduction}
Ontology systems usually provide reasoning and retrieval services that identify the basic facts that satisfy a requirement, and derive implicit knowledge using the ontology's inference axioms. In the context of the Semantic Web, the number of inferred facts can be extremely large. On one hand,
the amount of basic ontology facts (domain concepts and Web source annotations) can be considerable, and on the other hand,  {\it Open World} reasoning in Web ontologies may yield a large space of choices. Therefore, efficient evaluation strategies are needed in Web ontology's inference engines. 

In our approach, ontologies are formalized as a deductive database called a {\it Deductive Ontology Base} (DOB). The extensional database comprises
all the ontology language's statements 
that represent the explicit ontology knowledge. The intensional database corresponds to the set of deductive rules which define the semantics of the ontology language. 
We provide a cost-based optimization technique for Web ontologies represented as a DOB.

Traditional query optimization techniques for deductive databases systems include join-ordering strategies, and techniques that combine a bottom-up evaluation with top-down propagation of query variable bindings  in the spirit of the Magic-Sets algorithm \cite{ramakrishnan93survey}. Join-ordering strategies may be heuristic-based or cost-based; some cost-based approaches depend on the estimation of the {\it join selectivity}; others rely on the  {\it fan-out} of a literal \cite{conceptstore}.  Cost-based query optimization has been successfully used by relational database
management systems; however, these optimizers are not able to estimate the
cost or cardinality of data that do not exist a priori, which is the case of intensional predicates in a DOB.

We propose a hybrid cost model that combines two techniques for cardinality and cost estimation:
(1)  the sampling technique proposed in ~\cite{Lipton901,Lipton902} is applied for the estimation of the
evaluation cost and cardinality of intensional predicates, and (2) a  cost
model $\grave{a}$ la  System R cost model is used for the estimation of
the cost and cardinality of extensional predicates and the cost of conjunctive queries. 

Three evaluation strategies are considered for "joining" predicates in conjunctive queries. They are based on the Nested-Loop, Block Nested-Loop, and Hash Join operators of relational databases \cite{database03}.
To identify a good evaluation  plan,
we provide a dynamic-programming optimization algorithm that
orders subgoals in a query, considering estimates of the subgoal's evaluation cost.

We modeled a subset of the Web ontology language OWL Lite \cite{owl}  as a DOB, and performed experiments to study the predictive capacity of the cost model and the benefits of the ontology query optimization techniques. The study has been conducted over
synthetic and real-world OWL ontologies.
Preliminary results show that the cost-model estimates are pretty accurate and that optimized queries are significantly less expensive than non-optimized ones.

Our current formalism does not represent the OWL built-in constructor       
\textit{ComplementOf}. We stress that in practice this is not a severe limitation. For example, this operator is not used in any of the three real-world ontologies that we have         
studied in our experiments; and in the survey reported in \cite{ontnumber}, only 21 ontologies out of 688 contain this constructor.  

Our work differs from other systems in the Semantic Web that combine a Description Logics (DL) reasoner with a relational DBMS in order to solve the scalability problems for reasoning with individuals \cite{dllite,harslev04optimization,instancestore,dldb}. Clearly, all of these systems use the query optimization component embedded in the relational DBMS; however, they do not develop cost-based optimization for the implicit knowledge, that is, there is no estimation of the cost of data not known a priori. 

Other systems use Logic Programming (LP) to reason on large-scale ontologies. This is the case of the  projects described in \cite{grosof03dlp,motik06,motik03} . In  Description Logic Programs (DLP) \cite{grosof03dlp}, the expressive intersection between DL and LP without function symbols is defined. DL queries are reduced to LP queries and efficient LP algorithms are explored. The project described in \cite{motik06,motik03} reduces a $\mathcal{SHIQ}$ knowledge base to a Disjunctive Datalog program. Both projects apply Magic-Sets rewriting techniques but to the best of our knowledge, no cost-based optimization techniques have been developed. The OWL Lite$^{-}$ species of the OWL language proposed in \cite{bruijn04owllite} is based in the DLP project; it corresponds to the portion of the OWL Lite language that can be translated to Datalog. All of these systems develop LP reasoning with individuals, whereas in the DOB model we develop Datalog reasoning with both, domain concepts  and individuals. 

In \cite{eiter06}, an efficient bottom-up evaluation strategy for HEX-programs based on the theory of {\it splitting sets}  is described. In the context of the Semantic Web, these non-monotonic logic programs contain higher-order atoms and external atoms that may represent RDF and OWL knowledge. However, their approach does not include determining the best evaluation strategy according to a certain cost metric.

 

In the next section we describe our DOB formalism. Following this, we describe the DOB-S System architecture, Then, we model a subset of OWL Lite as a DOB and present a motivating example. Next, we develop our hybrid cost model and query optimization algorithm. We describe our experimental study and, finally, we point out our conclusions and future work.
\section{The Deductive Ontology Base (DOB)}
In general, an ontology knowledge base can be defined as:
\begin{Definition}[Ontology Knowledge Base]
An \textbf{ontology knowledge base O} is  a pair $O=\langle \mathcal{F},\mathcal{I}\rangle$, where 
$\mathcal{F}$ is a set of ontology facts that represent the explicit ontology structure (domain) and source annotations (individuals), and $\mathcal{I}$ is a set of axioms that allow the inference of new ontology facts regarding both domain and individuals. 
\end{Definition}

We will model $O$ as a deductive database which we call a {\it Deductive Ontology Base} (DOB). 
A DOB is composed of an Extensional Ontology 
Base (EOB) and an Intensional Ontology Base (IOB).  
Formally, a DOB is defined as:
\begin{Definition}[DOB]
Given an ontology knowledge base $O=\langle \mathcal{F},\mathcal{I}\rangle$, a \textbf{DOB} is a deductive database composed of a set of built-in EOB ground predicates representing  $\mathcal{F}$ and  a set of IOB built-in predicates representing $\mathcal{I}$, i.e. that define the semantics of the EOB built-in predicates.
\end{Definition}

The IOB predicate and DOB query definitions follow the Datalog language formalism \cite{Datalog}.
Next, we provide the definitions related to query-answering for DOBs.
\begin{Definition}[Valid Instantiation]
Given a Deductive Ontology Base $\mathcal{O}$, a set of constants $\mathcal{C}$ in $\mathcal{O}$, a set
of variables $\mathcal{V}$,  a rule $R$, and an interpretation $\mathbb{I}$ of $\mathcal{O}$ that  
corresponds to its Minimal Perfect Model \cite{Datalog}, a valuation \footnote{Given a set of variables $\mathcal{V}$ and a set of constants $\mathcal{C}$, a mapping or valuation $\gamma$ is a function $\gamma: \mathcal{V} \rightarrow \mathcal{C}$.} $\gamma$ is a \textbf{valid 
instantiation} of $R$ if and only if,  $\gamma(R)$ evaluates to true in
$\mathbb{I}$.
\label{vi}
\end{Definition}

\begin{Definition}[Intermediate Inferred Facts]
Given a Deductive Ontology Base $\mathcal{O}$, and a query 
$q:Q(\overline{X})\leftarrow \exists \overline{Y} B(\overline{X},\overline{Y}))$.
A proof tree for $q$ wrt $\mathcal{O}$ is defined as follows:   
\begin{itemize}
\item Each node in the tree is labeled by a predicate in $\mathcal{O}$.
\item Each leaf in the tree is labeled by a predicate in $\mathcal{O}$'s EOB.
\item The root of the tree is labeled by $Q$
\item For each internal node $N$ including the root, if $N$ is labeled
by a predicate $A$ defined by the rule $R$, $A(\overline{X})\leftarrow
\exists \overline{Y}
C(\overline{X},\overline{Y}))$, where $C(\overline{X},\overline{Y}))$ is
the conjunction of the predicates $C_1,...,C_n$, then, for each
valid instantiation of $R$, $\gamma$, the node $N$ has a sub-tree
whose root is  $\gamma(A(\overline{X}))$ and its children are 
respectively labeled $\gamma(C1)$,..., $\gamma(C_n)$.
\end{itemize}
The valuations needed to define all the valid instantiations in the
proof tree correspond to the {\bf Intermediate Inferred 
Facts} of $q$.
\label{defPT}
\end{Definition}

The number of  intermediate inferred facts measures the evaluation \textit{cost} of the query $Q$.
Additionally, since the valid instantiations of $Q$ in the proof tree correspond to the answers of the query, 
the \textit{cardinality} of $Q$ corresponds to the number of such instantiations.


%
%
%

Note that the sets of EOB and IOB built-in predicates of a DOB define an ontology framework, so  our model is not tied to any particular ontology language. To illustrate the use of our approach we focus on OWL Lite ontologies. 
\section{The DOB-S System's Architecture}
\begin{figure}[htb]
\centering
\includegraphics[width=0.7\linewidth]{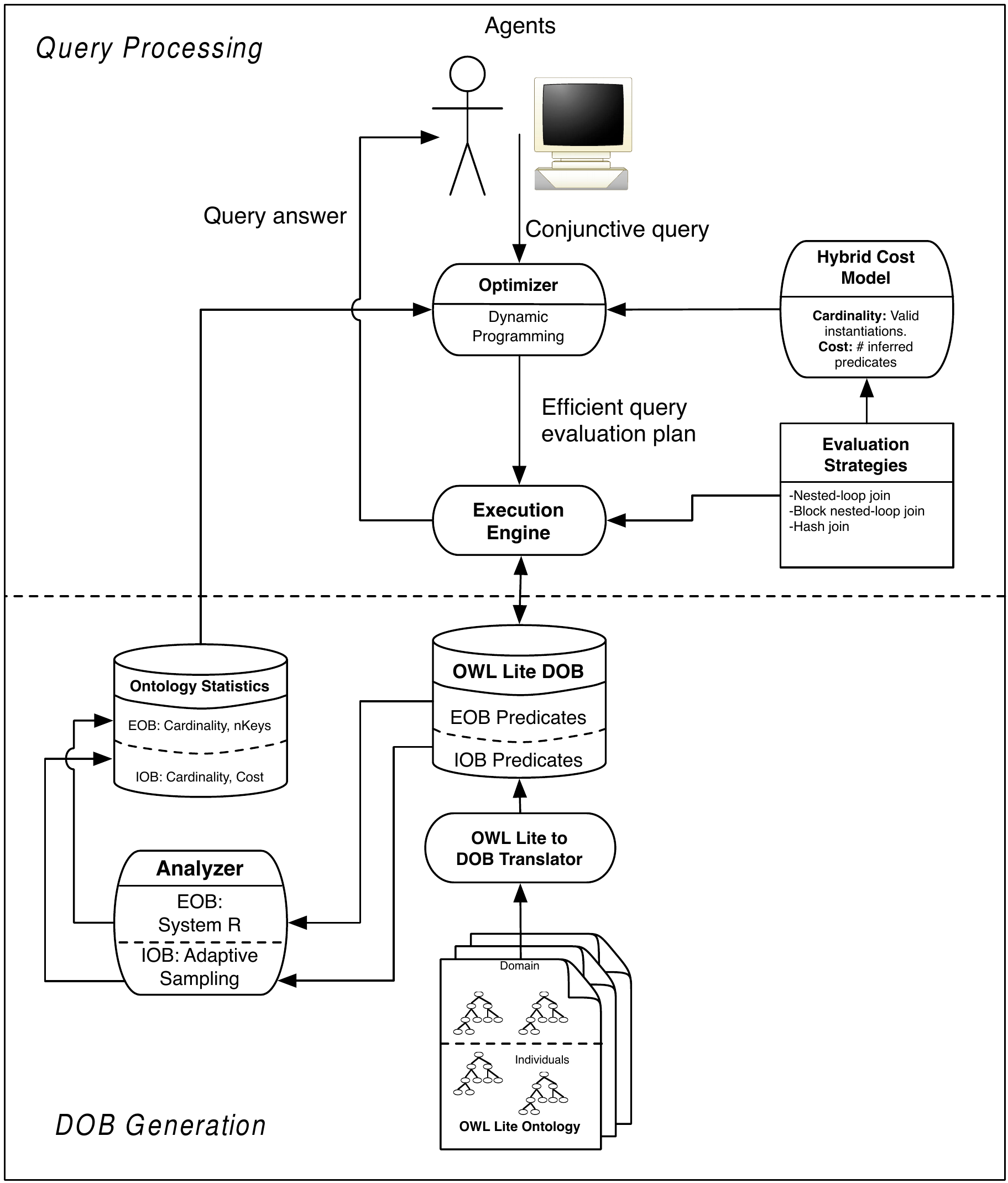}
\caption{DOB-S System Architecture}
\label{archit}
\end{figure}

DOB-S  is a system that allows an agent to pose efficient conjunctive queries against a set of ontologies. The system's architecture can be seen in Figure \ref{archit}. 

A subset of a given OWL ontology is translated into a DOB using an OWL Lite to DOB \textbf{translator}. EOB and IOB predicates are stored as a deductive database. Next, an \textbf{analyzer} generates the ontology's statistics: for each EOB predicate, the analyzer computes the number of facts or valid instantiations in the DOB (cardinality), and the number of different values for each of its arguments (nKeys); for each IOB predicate, an adaptive sampling algorithm \cite{Lipton901} is applied to compute cardinality and cost estimates. 

When an agent formulates a conjunctive query, the DOB-S system's \textbf{optimizer} generates an efficient query evaluation plan. A dynamic-programming optimizer is based in a \textbf{hybrid cost model}: it uses the ontology's EOB and IOB statistics, and estimates the cost of a query according the different evaluation strategies implemented. Finally, an \textbf{execution engine} evaluates the query plan and produces a query answer.

\section{OWL Lite DOB}


An OWL Lite ontology contains: (1) a set of axioms that provides information about classes and properties, and
(2) a set of facts that represents individuals  in the ontology, the classes they belong to, and the properties they participate in. 

Restrictions allow the construction of class definitions by restricting the values of their properties and their cardinality. Classes may also be defined through the intersection of other classes. Object properties represent binary relationships between individuals; datatype properties correspond to relationships between individuals and data values belonging to primitive datatypes. 

The subset of OWL Lite represented as a DOB does not include domain and range class intersection. 
Also, primitive datatypes are not handled; therefore, we do not represent ranges for Datatype properties\footnote{\begin{small}{\tt EquivalentClasses}\end{small}, \begin{small}{\tt EquivalentProperties}\end{small}, and \begin{small}{\tt allDifferent}\end{small} axioms, and the \begin{small}{\tt cardinality}\end{small} restriction are not represented because they are syntactic sugar for other language constructs.}. 
\subsection{OWL Lite DOB Syntax}
Our formalism, DOB,  provides a set of EOB built-in predicates that represents all the axioms and restrictions of an OWL Lite subset. 

EOB predicates are \textit{ground}, i.e., no variables are allowed as 
arguments. A set of IOB built-in predicates  represents the semantics of the EOB  predicates. We have followed the OWL Web Ontology Language Overview presented in \cite{owl}.

Table \ref{built1} illustrates the EOB and IOB built-in predicates for an OWL Lite subset\footnote{ We assume that the class {\it owl:Thing} is the default value for the domain and range of a property.}.
Note that some predicates refer to domain concepts (e.g. \begin{small}{\tt isClass, areClasses}\end{small}), and some to instance concepts (e.g. \begin{small}{\tt is isIndividual, areIndividuals}\end{small}).

\begin{table}[htbn]
\scriptsize
\centering
\caption{Some built-in EOB and IOB Predicates for a subset of OWL Lite} 
\begin{minipage}{\textwidth}
\begin{tabular}{l l}
\hline \hline \textbf{EOB PREDICATE} & \textbf{DESCRIPTION} \\ \hline
{\tt isOntology(O)} & An ontology has an Uri {\tt O}  \\ 
\noalign{\vspace{0.2cm}}
{\tt isImpOntology(O1,O2)} & Ontology {\tt O1} imports ontology {\tt O2} \\ 
\noalign{\vspace{0.2cm}}
{\tt isClass(C,O)} & {\tt C} is a class in ontology {\tt O}  \\ 
\noalign{\vspace{0.2cm}} 
{\tt isOProperty(P,D,R)} & {\tt P} is an object property with domain {\tt D} and range {\tt R}  \\ 
\noalign{\vspace{0.2cm}} 
{\tt isDProperty(P,D)} & {\tt P} is a datatype property with domain {\tt D}  \\ 
\noalign{\vspace{0.2cm}}  
{\tt isTransitive(P)} & {\tt P} is a transitive property \\ 
\noalign{\vspace{0.2cm}}  
{\tt subClassOf(C1,C2)} & {\tt C1} is subclass of {\tt C2}  \\ 
\noalign{\vspace{0.2cm}} 
{\tt AllValuesFrom(C,P,D)} & {\tt C} has property {\tt P} with  all values in {\tt D}   \\ 
\noalign{\vspace{0.2cm}} 
{\tt isIndividual(I,C)} & {\tt I} is an individual belonging to class {\tt C}  \\ 
\noalign{\vspace{0.2cm}}  
{\tt isStatement(I,P,J)} & {\tt I} is an individual that has property {\tt P} with value {\tt J} \\ \hline
 \textbf{IOB PREDICATE} & \textbf{DESCRIPTION} \\ \hline
{\tt areSubClasses(C1,C2)} & {\tt C1} are the direct and indirect subclasses of {C2}   \\ 
\noalign{\vspace{0.2cm}}  
¥{\tt areImpOntologies(O1,O2)} & {\tt O1} import the ontologies {\tt O2} directly and indirectly  \\ 
\noalign{\vspace{0.2cm}} 
{\tt areClasses(C,O)} & {\tt C} are all the classes of an ontology and its imported ontologies {\tt O}   \\ 
\noalign{\vspace{0.2cm}} 
{\tt areIndividuals(I,C)} & {\tt I} are the individuals of a class and all of its direct and indirect \\
¥  & superclasses {\tt C}; or \\ 
¥  & {\tt I} are the individuals that participate in a property and belong to  \\
¥  & its domain or range {\tt C}, or are values of a property with all values in {\tt C} \\ \hline \hline
\end{tabular}
\end{minipage}
\label{built1}
\end{table}

\begin{table}[htbn]
\scriptsize
\centering
\caption{Mapping OWL Lite subset to EOB Predicates} 
\begin{minipage}{\textwidth}
\begin{tabular}{ l l }
\hline \hline OWL ABSTRACT SYNTAX & EOB PREDICATES \\ \hline
$Ontology(O)$ & {\tt isOntology(O)} \\ 
\noalign{\vspace{0.2cm}}
$Individual(O1$ $value(owl:imports$ $O2))$ & {\tt impOntology(O1, O2)} \\ 
\noalign{\vspace{0.2cm}}
$Ontology(O),$ $Class(C$ $partial$ $Thing)$ & {\tt isClass(C,O)}  \\ 
\noalign{\vspace{0.2cm}}
$Class(A$ $partial$ $C)$ & {\tt subClassOf(A,C)} \\ 
\noalign{\vspace{0.2cm}}
$Class(C1$ $partial$ $restriction(P$ $allValuesFrom(C2)))$ & {\tt allValuesFrom(C1,P,C2)} \\ 
\noalign{\vspace{0.2cm}}
$Class(A$ $partial$ $C1\ldots Cn)$ & {\tt subClassOf(A,C1),...,} \\
¥ & {\tt subClassOf(A,Cn)}  \\ 
\noalign{\vspace{0.2cm}}
$ObjectProperty(P$ $domain(D))$,& {\tt isOProperty(P,D,R)} \\ 
$ObjectProperty(P$ $range(R))$  & ¥ \\ 
\noalign{\vspace{0.2cm}}
$DatatypeProperty(P$ $domain(D))$ & {\tt isDProperty(P,D)} \\ 
\noalign{\vspace{0.2cm}}
$Property(P$ $Transitive)$ & {\tt isTransitive(P)} \\ 
\noalign{\vspace{0.2cm}}
$Individual(I$ $type(C))$ & {\tt isIndividual(I,C)} \\ 
\noalign{\vspace{0.2cm}}
$Individual(I$ $value(P$ $J))$ & {\tt isStatement(I,P,J)} \\ \hline \hline
\end{tabular}
\end{minipage}
\label{built3}
\end{table}

\begin{table}[htbn]   
\scriptsize
\centering
\caption{Mapping OWL Lite subset Inference Rules to IOB Predicates} 
\begin{minipage}{\textwidth}
\begin{tabular}{ l l }
\hline \hline OWL LITE INFERENCE RULES &  IOB RULE DEFINITIONS \\ \hline
If {\tt subClassOf(C1,C2)} and {\tt subClassOf(C2,C3)} & {\tt areSubClasses(C1,C2):-subClassOf(C1,C2).} \\
then {\tt subClassOf(C1,C3)} ¥ & {\tt areSubClasses(C1,C2):-subClassOf(C1,C3),} \\
¥ & ~~~~~~~~~~~~~~~~~~~~~~~~~~~~~~~{\tt areSubClasses(C3,C2).} \\ 
\noalign{\vspace{0.2cm}}
If {\tt impOntology(O1,O2)} and {\tt impOntology(O2,O3)}  & {\tt areImpOntologies(O1,O2):-impOntology(O1,O2).} \\
 then {\tt impOntology(O1,O3)} & {\tt areImpOntologies(O1,O2):-impOntology(O1,O3),} \\ 
 ¥ &  ~~~~~~~~~~~~~~~~~~~~~~~~~~~~~~~~~~~{\tt areImpOntologies(O3,O2).} \\ 
 \noalign{\vspace{0.2cm}}
If {\tt isClass(C1,O2)} and {\tt impOntology(O1,O2)}  & {\tt areClasses(C,O):-isClass(C,O).} \\
then {\tt isClass(C1,O1)} & {\tt areClasses(C,O1):-isClass(C,O2),} \\
¥ &  ~~~~~~~~~~~~~~~~~~~~~~~~~{\tt areImpOntologies(O1,O2).} \\ 
\noalign{\vspace{0.2cm}}
If {\tt isSubClassOf(C1,C2)} and {\tt isIndividual(I,C1)} & {\tt areIndividuals(I,C):-isIndividual(I,C).} \\
then {\tt isIndividual(I,C2)} & {\tt areIndividuals(I,C2):-isIndividual(I,C1),} \\ 
¥ &  ~~~~~~~~~~~~~~~~~~~~~~~~~~~~~{\tt areSubClasses(C1,C2).} \\ 
If {\tt isStatement(I,P,J)} and {\tt isOProperty(P,C,R)} & {\tt areIndividuals(I,C):-isOProperty(P,C,R),} \\
then {\tt isIndividual(I,C)} &  ~~~~~~~~~~~~~~~~~~~~~~~~~~~~~{\tt areStatements(I,P,J).}  \\
If {\tt isStatement(I,P,J)} and {\tt isOProperty(P,D,C)} & {\tt areIndividuals(J,C): isOProperty(P,D,C),} \\
then {\tt isIndividual(J,C)} & ~~~~~~~~~~~~~~~~~~~~~~~~~~~~~{\tt areStatements(I,P,J).} \\
If {\tt isStatement(I,P,J)} and {\tt isDProperty(P,C)} & {\tt areIndividuals(I,C):-isDProperty(P,C),} \\
then {\tt isIndividual(I,C)} & ~~~~~~~~~~~~~~~~~~~~~~~~~~~~~{\tt areStatements(I,P,J).}  \\
If {\tt AllValues(C1,P,C)} and {\tt isStatement(I,P,J)} & {\tt areIndividuals(J,C):-isIndividual(I,C1),} \\
and {\tt isIndividual(I,C1)} then {\tt isIndividual(J,C)}  & ~~~~~~~~~~~~~~~~~~~~~~~~~~~~~{\tt allValuesFrom(C1,P,C),}  \\ 
¥  & ~~~~~~~~~~~~~~~~~~~~~~~~~~~~~{\tt areStatements(I,P,J).}  \\ \hline \hline
 \end{tabular}
\label{built2}
\end{minipage}
\end{table}
\subsection{OWL Lite DOB Semantics}
A model-theoretic semantics for an OWL Lite (subset) DOB is as follows:
\begin{Definition}[Interpretation]
An  \textbf{Interpretation} $I=(\Delta^I,\mathcal{P}^I,.^I)$  consists of:
\begin{itemize}
\item A non-empty interpretation domain $\Delta^I$  corresponding to the  union of the sets of valid URIs of ontologies, classes,  object and datatype properties, and individuals. These sets are pairwise disjoint.
\item A set of interpretations $\mathcal{P}^I$,  of the EOB and IOB built-in predicates in Table \ref{built1}.
\item An interpretation function $.^I$ which maps each n-ary built-in
predicate $p^I \in \mathcal{P}^I$ to an n-ary relation $\prod_{i=1}^{n} \Delta^I$. 
\end{itemize}
\end{Definition}
 \begin{Definition}[Satisfiability]
 Given an OWL Lite DOB $\mathcal{D}$, an interpretation $I$, and a predicate $p \in \mathcal{D}$, $I \models p$ iff:
 \begin{itemize}
 \item $p$ is an EOB predicate $p(t_1,...,t_n)$ and  $(t_1,...,t_n) \in p^I$.
 \item $p$ is an IOB predicate $R$:$H(\overline{X})\leftarrow \exists \overline{Y} B(\overline{X},\overline{Y})$, and whenever $I$ satisfies each predicate in the body $B$, $I$ also satisfies the predicate in the head $H$.
 \end{itemize}
 \end{Definition}
\begin{Definition}[Model]
Given an OWL Lite DOB $\mathcal{D}$ and an interpretation $I$, $I$ is a \textbf{model} of $\mathcal{D}$ iff for every predicate $p \in \mathcal{D}$, $I  \models p$.
\end{Definition}
\subsection{Translation of OWL Lite to OWL Lite DOB}
A definition of a translation map from OWL Lite to OWL Lite DOB is the following:
\begin{Definition}[Translation]
Given an OWL Lite  theory $\mathcal{O}$ and an OWL Lite DOB theory $\mathcal{D}$, an \textbf{OWL Lite to DOB Translation} $\mathcal{T}$ is a  function $\mathcal{T}: \mathcal{O} \rightarrow \mathcal{D}$.
\end{Definition}
Given an OWL Lite  ontology $\mathcal{O}$, an OWL Lite DOB ontology $\mathcal{D}$  is defined as follows:
\begin{itemize}
\item{(Base Case) If  $o$ is an axiom or fact belonging to the sets of axioms or facts of $\mathcal{O}$, then an EOB predicate $\mathcal{T}(o)$ is defined according to the EOB mappings in Table \ref{built3}}.
\item{If $o$ is an OWL Lite inference rule, then an IOB predicate $\mathcal{T}(o)$ is defined according to the IOB mappings  in Table \ref{built2}}. 
\end{itemize}
The translation ensures that the following theorem holds:
\begin{Theorem}
Let $\mathcal{O}$ and  $\mathcal{D}$ be OWL Lite  and OWL Lite DOB theories respectively, and $\mathcal{T}$ be an OWL Lite to DOB Translation such that, $\mathcal{T}(\mathcal{O})=\mathcal{D}$, then $\mathcal{D} \models \mathcal{O}$.      
\end{Theorem} 
\section{A Motivating Example}

Consider  a 'cars and dealers' domain ontology {\tt carsOnt} and Web source ontologies {\tt source1} and {\tt source2}. Source {\tt source1} publishes information about all types of vehicles and dealers, whereas 
{\tt source2} is specialized in SUVs. 

The OWL Lite ontologies can be seen in Table \ref{exowllite}.
 \begin{table}[htbn]
\scriptsize
\centering
\caption{Example OWL Lite ontology} 
\begin{minipage}{\textwidth}
\begin{tabular}{ l l l }
\hline \hline \textbf{Ontology carsOnt} & \textbf{Ontology source1} & \textbf{Ontology source2}\\ \hline
{\tt Class (vehicle partial Thing)} & {\tt imports carsOnt} & {\tt imports carsOnt}  \\ 
\noalign{\vspace{0.2cm}}
{\tt Class (suv partial vehicle)} & ¥ & {\tt individual(s123 type(suv))}   \\ 
\noalign{\vspace{0.2cm}} 
{\tt Class (car partial vehicle)}  & ¥ & ¥ \\ 
\noalign{\vspace{0.2cm}}
{\tt DataProperty(price domain(vehicle))}  & ¥ & ¥  \\ 
\noalign{\vspace{0.2cm}} 
{\tt Class (dealer partial Thing)} & ¥ & ¥ \\ 
\noalign{\vspace{0.2cm}} 
{\tt ObjectProperty(sells domain(dealer))} & ¥ & ¥ \\ 
\noalign{\vspace{0.2cm}} 
{\tt ObjectProperty(sells range(vehicle))} & ¥ & ¥ \\ 
\noalign{\vspace{0.2cm}}
{\tt DataProperty(traction domain(suv))} & ¥ &¥ \\ 
\noalign{\vspace{0.2cm}}
{\tt DataProperty(model domain(vahicle))} & ¥ & ¥ \\ \hline \hline
\end{tabular}
\end{minipage}
\label{exowllite}
\end{table}

 A portion of the example's EOB can be seen in Table \ref{example}.
\begin{table}[htbn]
\scriptsize
\centering
\caption{Example DOB ontology}
\begin{minipage}{\textwidth}
\begin{tabular}{ l l l }
\hline \hline \textbf{EOB PREDICATES}  \\ \hline
{\tt isOntology(carsOnt)} & {\tt isOntology(source1)} & {\tt isOntology(source2)}  \\ 
\noalign{\vspace{0.2cm}}
{\tt impOntology(source1,carsOnt)} & {\tt impOntology(source2,carsOnt)} & {\tt isClass(vehicle,carsOnt)}   \\ 
\noalign{\vspace{0.2cm}}
{\tt isClass(vehicle,carsOnt)} & {\tt isClass(dealer,carsOnt)} & {\tt subClassOf(car,vehicle)} \\ 
\noalign{\vspace{0.2cm}} 
{\tt subClassOf(suv,vehicle)}  & {\tt isOProperty(sells,dealer,vehicle)} & {\tt isDProperty(model,vehicle)}  \\ 
\noalign{\vspace{0.2cm}} 
{\tt isDProperty(price,vehicle)} & {\tt isDProperty(traction,suv)} & {\tt isIndividual(s123,suv)} \\ \hline \hline
\end{tabular} 
\end{minipage}
\label{example}
\end{table}

To illustrate a rule evaluation, we will take a query \begin{small}{\tt q}\end{small} that asks  for {\it the Web sources that publish information about 'traction'}:
\begin{center}
\begin{small}
{\tt
q(O):-areClasses(C,O),isDProperty(traction,C).}
\end{small}  
\end{center}
The answer to this query corresponds to all the ontologies with classes characterized by the property \begin{small}{\tt traction}\end{small}, i.e., ontologies {\tt source1}, 
{\tt source2} and {\tt carsOnt}. 

If we invert the ordering of the first two predicates in \begin{small}{\tt q}\end{small}, we will have an equivalent query \begin{small}{\tt q'}\end{small}:
\begin{center}
\begin{small}
{\tt
q'(O):-isDProperty(traction,C),areClasses(C,O).}
\end{small}
\end{center}

The cost or total number of inferred facts for \begin{small}{\tt q}\end{small} is larger than the cost for   \begin{small}{\tt q'}\end{small}. In \begin{small}{\tt q}\end{small},  the number of instantiations or cardinality for the first intensional predicate \begin{small}{\tt areClasses(C,O)}\end{small} is twelve, four for each ontology, as \begin{small}{\tt source1}\end{small} and  \begin{small}{\tt source2}\end{small} inherit the classes in  \begin{small}{\tt carsOnt}\end{small}. 
The cost of inferring these facts  is dependent on the cost of evaluating the \begin{small}{\tt areClasses}\end{small} rule. In \begin{small}{\tt q'}\end{small}, for the first subgoal \begin{small}{\tt isDProperty(traction,C)}\end{small}, we have one instantiation: \begin{small}{\tt isDProperty(traction,suv)}\end{small}. Again, the cost of inferring this fact depends on the cost of the
 \begin{small}{\tt isDProperty}\end{small} predicate.
 
Note that statistics on the size and argument values of the EOB \begin{small}{\tt isDProperty}\end{small} predicate can be computed, whereas statistics for the IOB \begin{small}{\tt areClasses}\end{small} predicate will have to be estimated as data is not known a priori. Once the cost of each query predicate is determined, we may apply a cost-based join-ordering optimization strategy.
\section{DOB Hybrid Cost Model}

The process of answering a query relies on inferring facts from the predicates in the DOB. 
Our cost metric is focused on the number of intermediate facts that need to be inferred in 
order to answer the query. The objective is to find an order of the predicates in the body
of the query, such that the number of intermediate inferred facts is reduced.
We will apply a join-ordering optimization strategy $\grave{a}$ la System R  using 
Datalog-relational equivalences \cite{Datalog}.  To estimate the cardinality and evaluation cost of the intensional predicates, we have applied an adaptive sampling technique. Thus, we propose a  hybrid cost model which combines 
adaptive sampling and traditional relational cost models.
\subsection{Adaptive Sampling Technique}
We have developed a
sampling technique that is based on the {\em adaptive sampling method}
proposed by Lipton, Naughton, and
Schneider~\cite{Lipton901,Lipton902}.  This technique assumes that there is a
population $\mathbb{P}$ of all the different valid instantiations of a predicate $P$, and that $\mathbb{P}$ is divided into $n$ partitions according to the $n$ possible instantiations of one or more arguments of $P$.
 Each element in $\mathbb{P}$  is related to its evaluation cost and cardinality, and the population
$\mathbb{P}$  is characterized  by the statistics mean and variance.

The objective of the sampling  is
to identify a sample of the population $\mathbb{P}$, called
$\mathbb{EP}$, such that  the mean and variance of the cardinality (resp. evaluation cost)
 of $\mathbb{EP}$ are valid to within a predetermined
accuracy and confidence level.  

To estimate the mean of the cardinality
(resp. cost) of  $\mathbb{EP}$, say $\overline{Y}$, within $\frac{\overline{Y}}{d}$ with probability
$p$, where $0 \leq p<1$ and $d>0$, the sampling method assumes an {\it urn} model.

The urn has $n$ balls from which $m$ samplings are repeatedly
taken, until the sum $z$ of the cardinalities (resp.  costs)  of the
samples is greater than $\alpha\times (\frac{S}{Y})$, where $\alpha=\frac{d\times (d+1)}{(1-\sqrt{p)}}$.
The estimated mean  of the cardinality (resp. cost) is: $\overline{Y}=\frac{z}{m}$.

The values $d$ and $\frac{1}{(1-\sqrt{p)}}$ are associated with the relative error and the
confidence level, and $S$ and $Y$ represent the cardinality
(resp. cost) variance and mean of $\mathbb{P}$. 
Since statistics of
$\mathbb{P}$ are unknown, the upper bound $\alpha \times \frac{S}{Y}$ is replaced by $\alpha \times b(n)$.

To approximate $b(n)$ for cost and cardinality estimates, we apply Double Sampling \cite{Ling92}. In the first stage
 we randomly evaluate $k$ samples and take the maximum value among them:
 
 \begin{small}$b(n)=max_{i=1}^{k}(card(P_i))$ (resp. $b(n)=max_{i=1}^{k}(cost(P_i)))$, where $1\leq k\leq n$\end{small} \\ 
 
It has been shown that a few samples are necessary in order for the distribution of the sum to begin to look normal. Thus, the factor $1/(1-\sqrt{p})$ may be improved by central limit theorem \cite{Lipton902}. This improvement allows us to achieve accurate estimations and lower bounds.




\subsubsection{Estimating cardinality.} 
Given an intensional predicate $P$, the {\it cardinality} of $P$ corresponds to the
number  of  valid instantiations of $P$ (Definition ~\ref{vi}).
In our previous example, the number of ontology values obtained in the answer of the query  is estimated using this metric.

To estimate the cardinality of $P$, we execute the adaptive sampling algorithm explained before, by selecting any argument of $P$, and partitioning $\mathbb{P}$  according to the chosen argument. The cardinality estimation will be \begin{small}$card(P) = \overline{Y}\times n$\end{small}, where $n$ is the number of partitions, i.e. the number of different instantiations for the chosen argument.

Note that once the cardinality of the non-instantiated $P$ is estimated, we can estimate the cardinality of the instantiated predicate by using the selectivity value(s) of the instantiated argument(s).
\subsubsection{Estimating cost.}
The {\it cost} of $P$ measures the number of intermediate inferred facts (Definition ~\ref{defPT}).
For instance, to estimate the cost of a predicate \begin{small}$P(X,Y)$\end{small},  we consider the different instantiation patterns that the predicate can have, i.e., we independently estimate the cost  
for \begin{small}$P(X^{b},Y^{b})$, $P(X^{b},Y^{f})$, $P(X^{f},Y^{b})$\end{small} and \begin{small} $P(X^{f},Y^{f})$\end{small},
 where {\it b} and {\it f} indicate that the argument is bound and free, respectively.
 
The computation of several cost estimates is necessary because in Datalog top-down evaluation \cite{Datalog}, the cost of an instantiated
intensional predicate cannot be accurately estimated from the cost of a non-instantiated
predicate (using selectivity values). Instantiated arguments will propagate in the IOB rule's body through sideways-passing, and cost varies according to the binding patterns.
For example,  the cost of \begin{small}{\tt areClasses(C1$^b$,C2$^f$)}\end{small} may be 
smaller than the cost of \begin{small}{\tt areClasses(C1$^f$,C2$^b$)}\end{small}, i.e., the bound argument \begin{small}{\tt C1}\end{small} "pushes" instantiations in the definition of the rule:

\begin{small}{\tt areSubClasses(C1,C2):-isSubClass(C1,C3),areSubClasses(C3,C2).}\end{small}

making its body predicates more selective.


For \begin{small}$P(X^{b},Y^{b})$, $P(X^{b},Y^{f})$\end{small} and \begin{small}$P(X^{f},Y^{b})$\end{small}, we partition $\mathbb{P}$ according to the bound arguments. In these cases we are estimating the cost of one partition. Therefore,  \begin{small}$cost(P) = \frac{\overline{Y}\times n}{n}=\overline{Y}$\end{small}.

Finally, to estimate the cost of \begin{small}$P(X^{f},Y^{f})$\end{small}, we choose an argument of $P$ and
partition $P$ according to the chosen argument. To reduce the cost of computing the estimate, we choose the most selective argument. The cost estimate is \begin{small}$cost(P) =
\overline{Y}\times n$\end{small}.

\subsubsection{Determining the number of partitions $n$.}
For both, cost and cardinality estimates, we need to determine the number of possible instantiations, $n$, of the chosen argument. This value depends on the semantics of the particular predicate. For instance, for an interpretation $I$, \begin{small}\[areClasses(Class,Ont)^I \subseteq \mathcal{C} \times \mathcal{O}\]\end{small} where $\mathcal{C}$ is the set of valid class URIs and $\mathcal{O}$ is the set of valid ontology URIs. $|\mathcal{C}|$ corresponds to the number of EOB predicates \begin{small}$isClass(Class,Ont)$\end{small}, i.e., \begin{small}\[|\mathcal{C}|=Card(isClass(Class,Ont))\]\end{small}Similarly,  \begin{small}$|\mathcal{O}|$=$Card(isOntology(Ont))$\end{small}; these cardinalities  are  pre-computed offline. We assume that the values are uniformly distributed.

\subsection{System R Technique}
To estimate the cardinality and cost of two or more predicates, we use the cost model proposed 
in System R. 
The cardinality of the conjunction of predicates  \begin{small}$P_1$,$P_2$\end{small} is described by the following
 expression:
\begin{small}
\[card(P_1, P_2) = card(P_1)\times card(P_2) \times
 reductionFactor(P_1,P_2)\]                      
\end{small} 
\begin{small}$reductionFactor(P_1,P_2)$\end{small} reflects the impact of the sideways passing variables in reducing the cardinality of the result. This value is computed assuming that sideways passing variables are independent and each is uniformly distributed \cite{SystemR}.  
For cost estimation, we consider three evaluation strategies:
\begin{enumerate}
\item{Nested-Loop Join}\\
Following a Nested-Loop Join evaluation strategy, for each valid instantiation in $P_1$, we retrieve a valid instantiation in $P_2$ with a matching "join" argument value:

\begin{small}
\[cost(P_1,P_2) = cost(P_1)+card(P_1) \times cost^{inst}(P_2)\]
\end{small}
\begin{small}$cost^{inst}(P_2)$\end{small} corresponds to the estimate of the cost of the predicate $P_2$  where the "join" arguments are instantiated in $P_2$, i.e., all the sideways passing variables from $P_1$ to $P_2$ are bound in $P_2$. 
These binding patterns were considered during the sampling-based estimation of the cost of $P_2$. 
\item{Block Nested-Loop Join}\\
Predicate $P_1$ is evaluated into blocks of fixed size, and then each block is "joined" with $P_2$.
\begin{small}
\[cost(P_1,P_2) = cost(P_1)+ \lceil \frac{card(P_1)}{BlockSize} \rceil \times cost(P_2)\]
\end{small}
\item{Hash Join}\\
A hash table is built for each predicate according to their join argument. The valid instantiations of predicates $P_1$ and $P_2$ with the same hash key will be joined together:
\[cost(P_1,P_2) = cost(P_1)+cost(P_2)\]
\end{enumerate}
 

Although the sampling technique is appropiate for estimating a 
single predicate, it may be 
inefficient for estimating the size of a conjunction of more than two predicates. 

The sampling algorithm in  \cite{Lipton901} suggests that for a conjunction of two predicates, $P,Q$, if  the size of $P$ is $n$, the query is n-partitionable, i.e., for each valid instantiation $p$ in $P$, the corresponding partition of $Q$ contains all the valid instantiations $q$ in $Q$ such that $q$ "joins" $p$.
Therefore, when the size of the first predicate in a query is small, its sample size may be larger. This problem can be extended to conjunctive queries with several subgoals, so when the number of intermediate results is small, sampling time may be as large as evaluation time. 

%

\subsection{Query Optimization}
In Figure ~\ref{algo}, we present the algorithm used to optimize the body
of a query. 
The proposed
optimization algorithm extends the System R dynamic-programming algorithm by identifying orderings of the $n$
EOB and IOB predicates in a query. During each iteration of the algorithm, the best intermediate 
sub-plans are chosen based on  cost and cardinality. In the last iteration, final plans are constructed and the best plan is selected in terms of the cost metric.

During each iteration {\it i} between 2 and {\it n-1},  different orderings
of the predicates are analyzed. Two
subplans are considered equivalents if and only if, they are composed by the same
predicates. 
A subplan $SP_i$ is better than a subplan $SP_j$
if and only if, the cost and cardinality of $SP_j$ are greater than the cost
and cardinality of $SP_i$, respectively. If $SP_i$ cost is greater than $SP_j$ cost, but
$SP_j$ cardinality is greater than $SP_i$ cardinality, i.e. they
are un-comparable, then the equivalence class is annotated with
the two subplans. 

\begin{table}
{\scriptsize
\centering
\caption{Query Optimization Algorithm}
\begin{minipage}{\textwidth}
\begin{tabular}{p{4.75in}}
\hline \hline \textbf{Algorithm Dynamic Programming}  \\ \hline
{\it INPUT}: {\tt Predicate}: a set of predicates, $P_1$,...,$P_n$.
{\it OUTPUT}: {\tt OrderedPredicate}: an ordering of {\tt Predicate}
\begin{enumerate}
\item {\tt SubPaths}={\tt Predicate};
\item For i=1 to n
\begin{enumerate}
\item For each solution $Sub_j$ in {\tt SubPaths}
\begin{enumerate}
\item For each predicate $P_z$ in {\tt Predicate}
\begin{itemize}
\item If there are sideways passing variables from $Sub_j$ to $P_z$, 

then add $Sub$= $Sub_j$,$P_z$ to {\tt NewSubPaths}
\end{itemize}
\end{enumerate}
\item Remove from {\tt NewSubPaths} any subpath $Sub_k$ iff there is another
subpath $Sub_l$ in {\tt NewSubPaths}, such that, $Sub_l$ and $Sub_k$ are
{\bf equivalent}, and  $Sub_l$ is {\bf better} than $Sub_k$.
\item {\tt SubPaths}={\tt NewSubPaths}
\item Reset {\tt NewSubPaths}
\end{enumerate}
\item Return the path in {\tt SubPaths} with lowest cost.
\end{enumerate}
\\ \hline
\end{tabular}
\end{minipage}
}
\label{algo}
\end{table}

%
%

\section{Experimental Results}
An experimental study was conducted for synthetic and real-world ontologies. Experiments on synthetic ontologies were executed on a SunBlade 150 (650MHz) with 1GB RAM; experiments on real-world ontologies were executed on a SunFire V440 (1281MHz) with 16GB RAM. Our system was implemented in SWI-Prolog 5.6.1.

We have studied three real-world ontologies: Travel  \cite{schemaweb}, EHR\_RM \cite{protege}, and GALEN \cite{galen}. 

Our cost metrics are the number of intermediate facts  for synthetic and real-world ontologies,  and the evaluation time for real-world ontologies. In our experiments, the sampling  parameters  $d$ (the error), $p$ (the confidence level), and $k$ (the size of the sample for the first stage) were set to 0.2, 0.7 and 7, respectively. 
We developed two sets of experiments according to the evaluation strategies considered: (1) the Nested-Loop join evaluation strategy, and (2) the combination of Nested-Loop, Block Nested-Loop and Hash join evaluation strategies.
Our study consisted of the following:
 \begin{itemize}
 \item {\it Cost Model Predictive Capability}:  In Figure \ref{combinatedFigure1}a, we report the correlation among the estimated values and the actual cost for synthetic ontologies considering the Nested-loop Join evaluation strategy. Synthetic ontologies were randomly generated following a uniform distribution. We generated ten ontology documents and three chain and star queries with three subgoals for each ontology; the cost of each ordering was estimated with our cost model, and each ordering was then evaluated against the ontology; this gives us a total of six hundred queries. The correlation is 0.92. 

\begin{figure}[h]
\centering
\subfigure[]{\includegraphics[width= 0.45\linewidth]{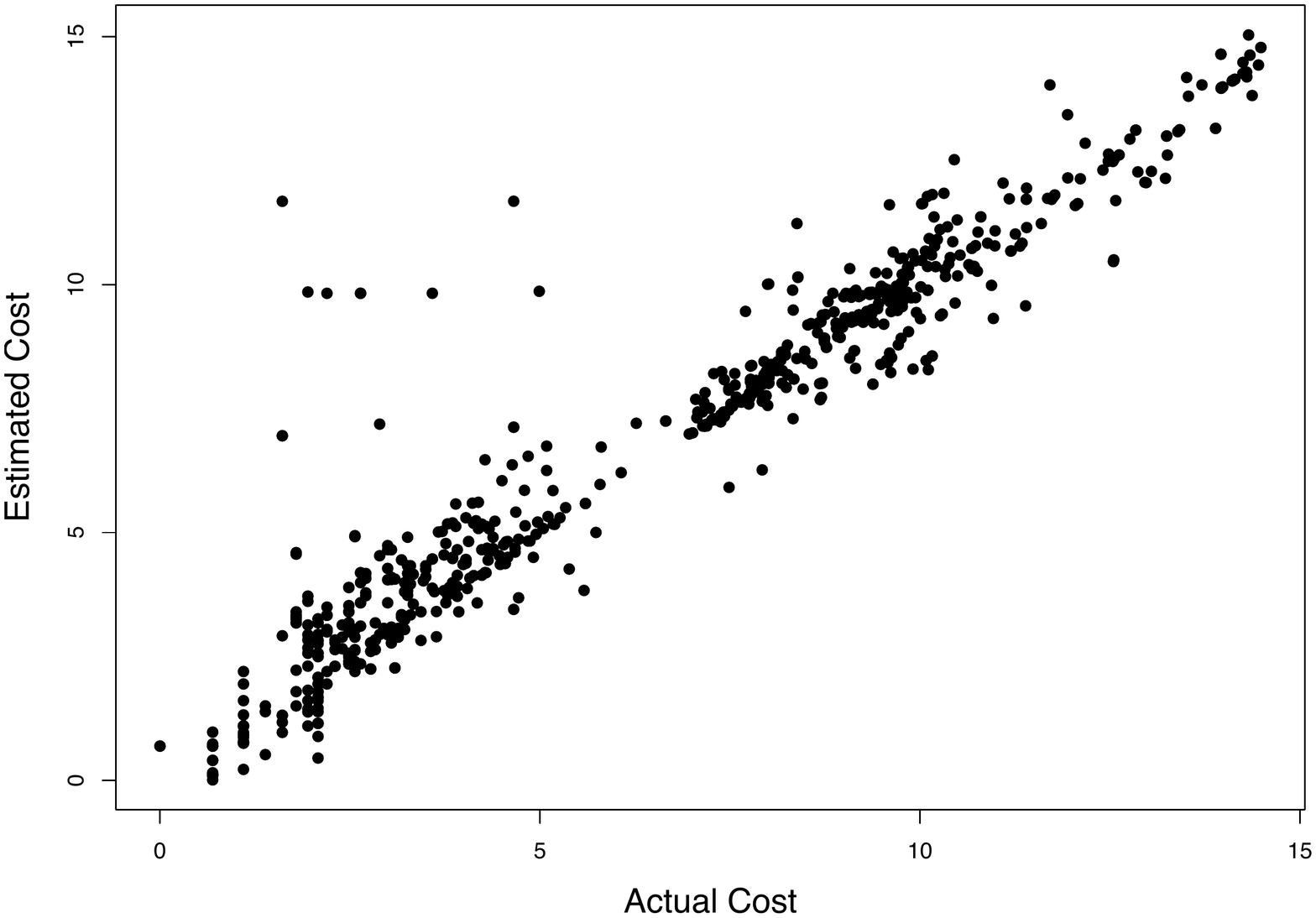}}\qquad
\subfigure[]{\includegraphics[width=0.45\linewidth]{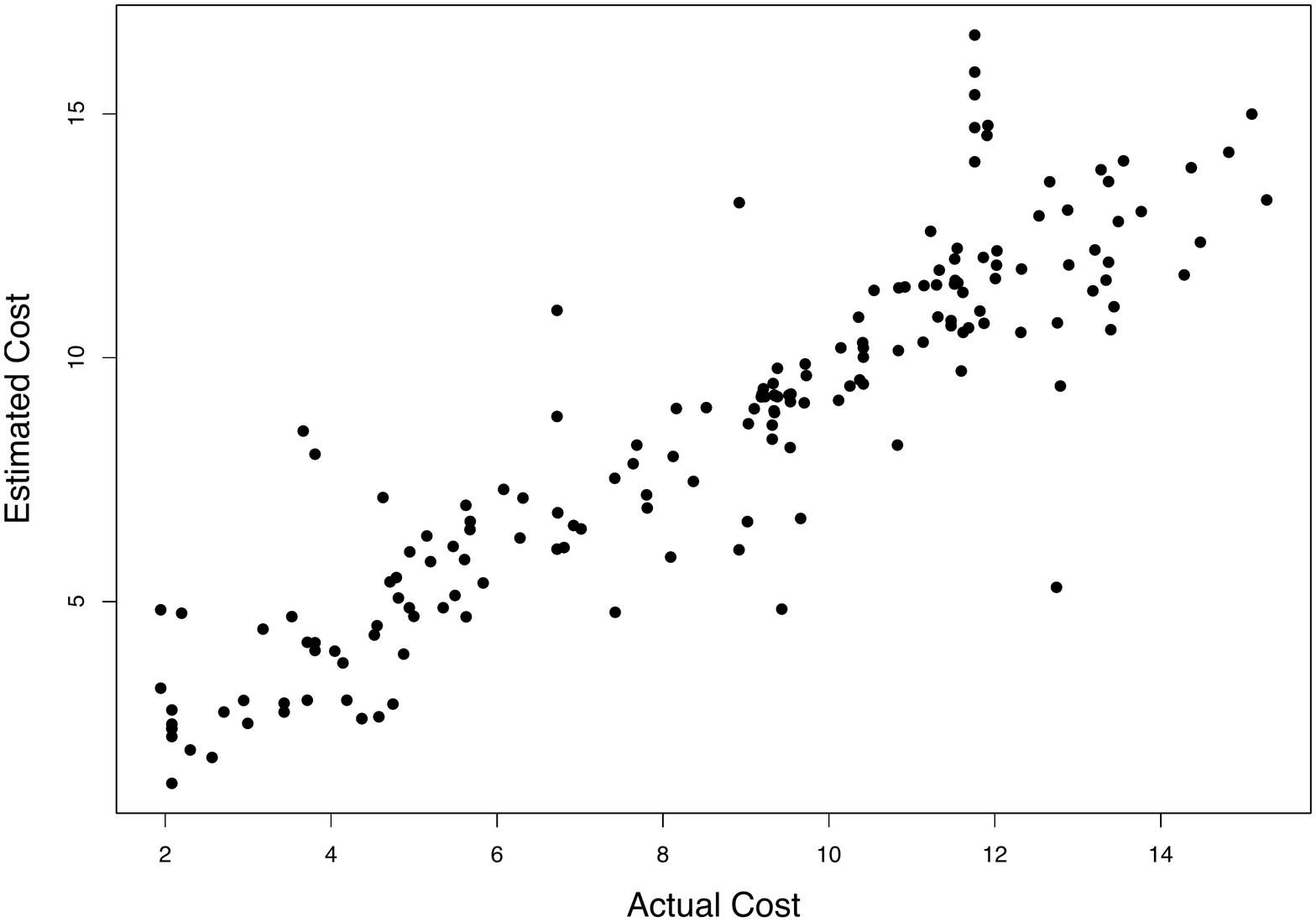}}
\caption{(a) Correlation of estimated cost to actual cost (log. scale) - nested-loop join - 
Synt. ontologies; (b) Correlation of estimated cost to actual cost (log. 
scale) - nested-loop join - GALEN}
\label{combinatedFigure1}
\end{figure}


\begin{figure}[htb]
\subfigure[]{\includegraphics[width=0.45\linewidth,height=2.20in]{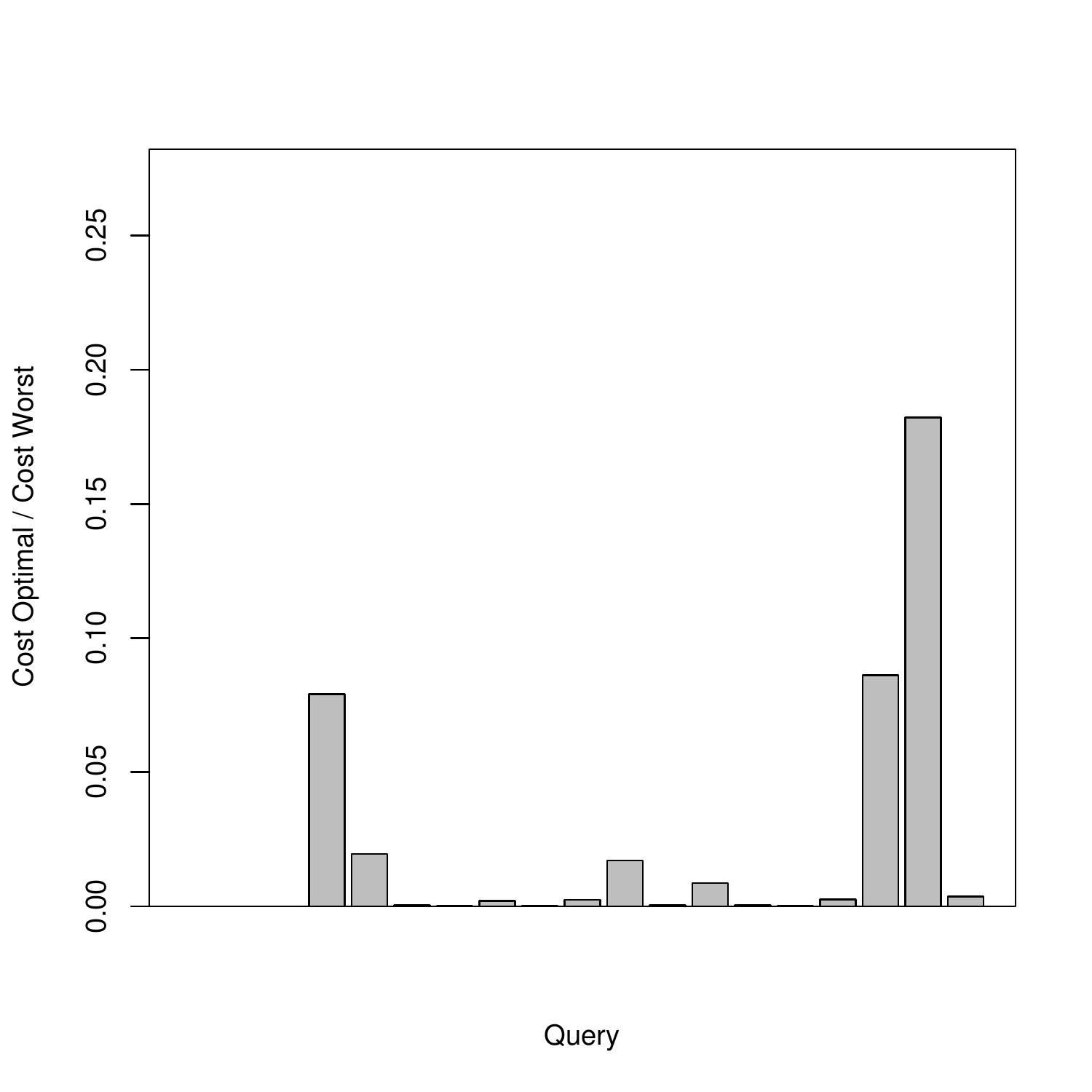}}\qquad
\subfigure[]{\includegraphics[width=0.45\linewidth,height=2.20in]{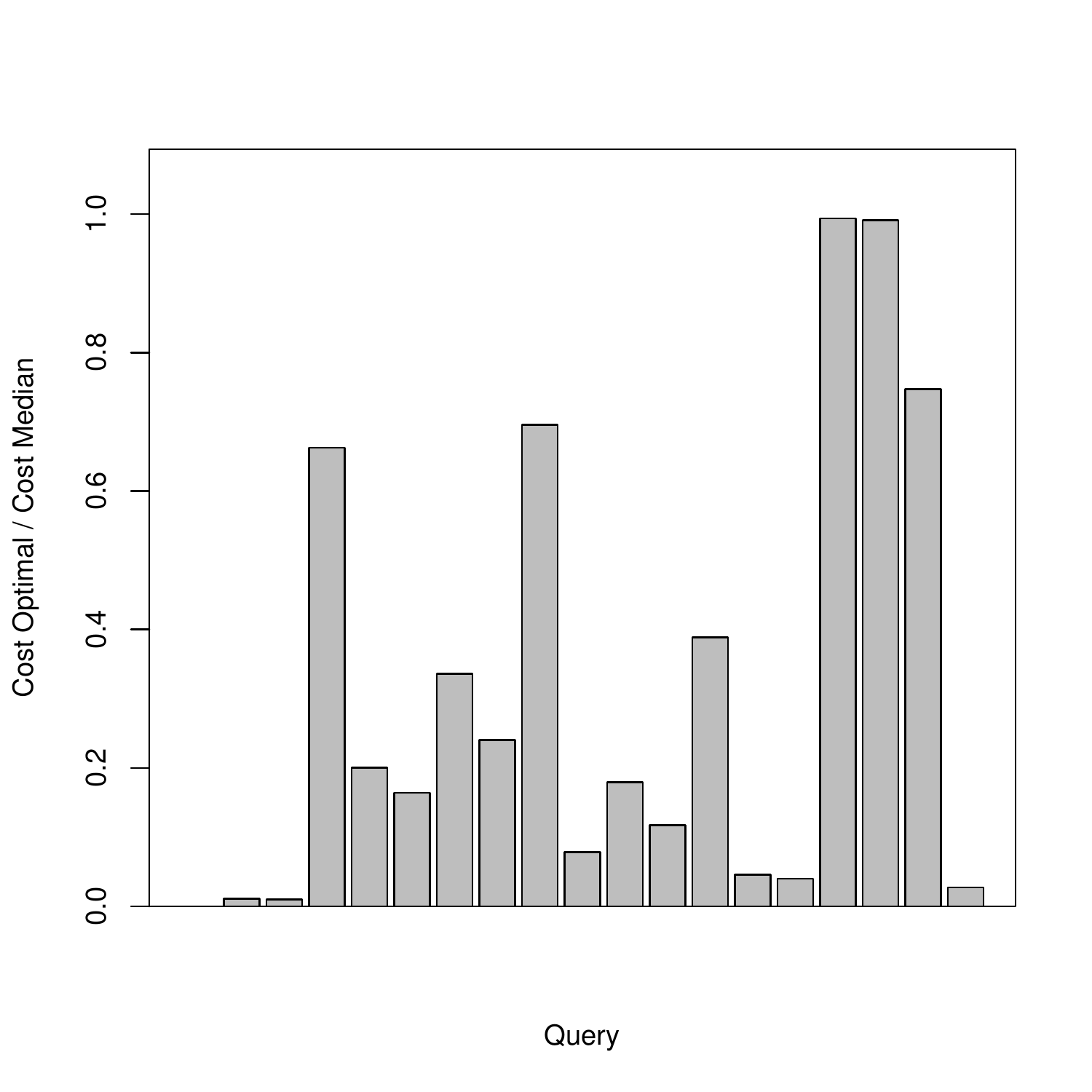}}
\caption{(a) \#Pred. optimal ordering vs. \#Pred. worst ordering - nested-loop-join - 
Synt. Ontologies; (b) \#Pred. optimal ordering vs. \#Pred. median ordering - nested-loop-join - 
Synt. Ontologies} 
\label{combinatedFigure2}
\end{figure}

\begin{figure}[htb]
\subfigure[]{\includegraphics[width=0.45\linewidth,height=2.20in]{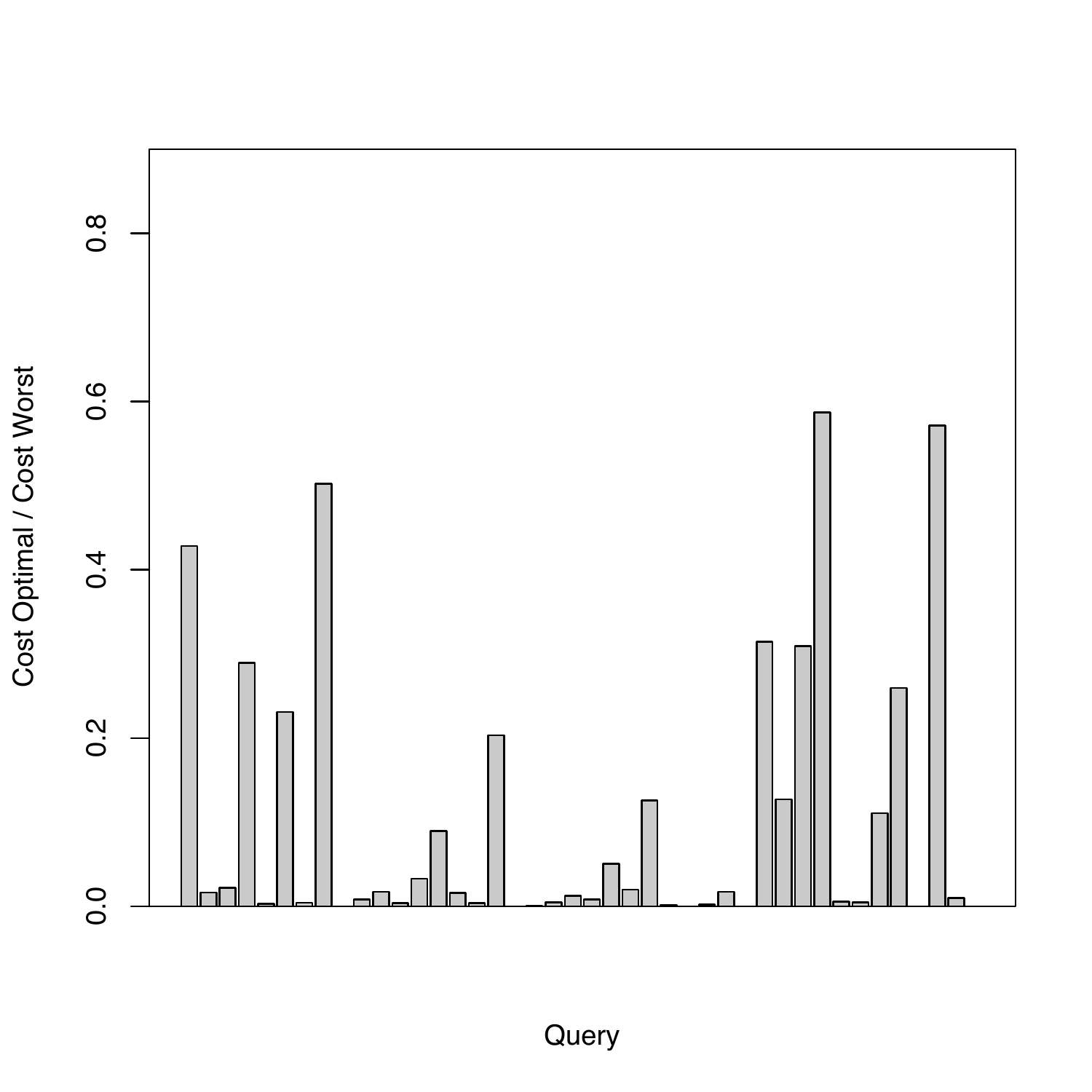}}\qquad
\subfigure[]{\includegraphics[width=0.45\linewidth,height=2.20in]{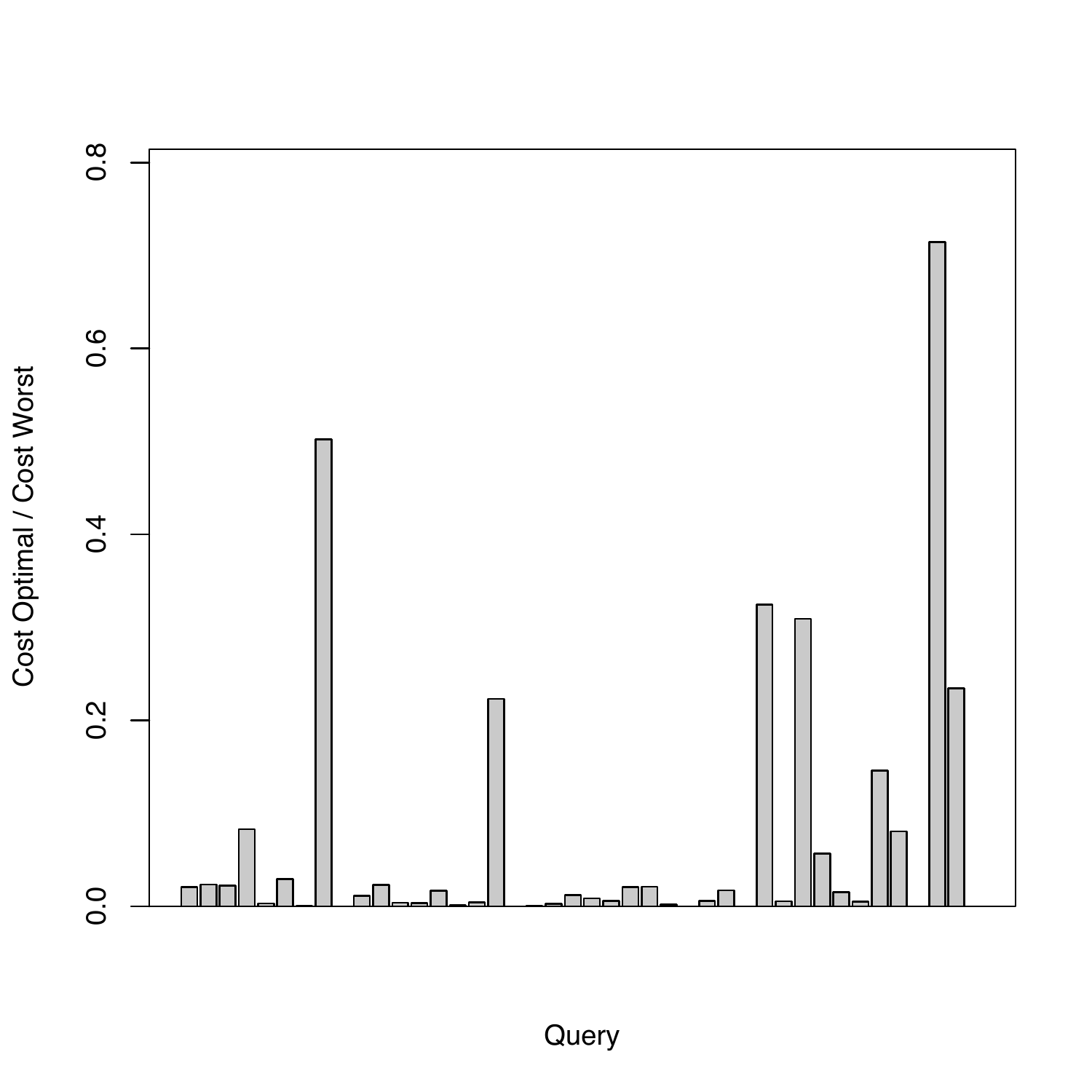}}
\caption{(a) \#Pred. optimal ordering vs. \#Pred. worst ordering - nested-loop-join - EHR\_RM; (b)  \#Pred. optimal ordering vs. \#Pred. worst ordering 
- combination evaluation strategies - EHR\_RM} 
\label{combinatedFigure3}
\end{figure}

\begin{table}[htbn]
\scriptsize
\centering
\caption{Correlation values for real-world ontologies}
\begin{minipage}{\textwidth}
\begin{tabular}{ l c  c}
\hline \hline 
¥ & \textbf{Nested-Loop Join} & \textbf{Three Evaluation Strategies} \\ \hline
\textbf{Travel} & 0.96 & 0.94  \\ 
\textbf{EHR\_RM} & 0.98 & 0.92 \\
\hline \hline
\end{tabular} 
\end{minipage}
\label{correlation}
\end{table}

 In Figure \ref{combinatedFigure1}b, 
 we report the same correlation metric for the real-world ontology GALEN, and the value is 0.62. 
In Table \ref{correlation}, we present correlation values for the real-world ontologies Travel and EHR\_RM for our two sets of experiments: the accuracy of the Nested-Loop join cost model is similar to the accuracy of the cost model that considers the combination of the three evaluation strategies.
\item {\it Cost improvements}: We also conducted experiments to study cost improvement using the optimizer. We evaluated all the orderings of each query, then we ran the optimizer and evaluated the optimized query. Figure \ref{combinatedFigure2}a reports  the ratio of the cost of the optimal ordering to the cost of the worst ordering considering only nested-loop join, 
 $\frac{costOptimalOrdering}{costWorstOrdering}$, for queries against synthetic ontologies.  For synthetic  ontologies, this ratio is less than 10\% for most of the queries. We also computed the proportion of the optimal ordering cost with respect to the median ordering cost. The results for synthetic ontologies show that the optimal ordering cost is less than 40\% of the median for fifteen of twenty queries; this result can be observed in Figure \ref{combinatedFigure2}b.
 
In Figure  \ref{combinatedFigure3}a, we report   the ratio of the cost of the optimal ordering to the cost of the worst ordering considering only nested-loop join for EHR\_RM. Additionally, Figure ~\ref{combinatedFigure3}b reports the same metric considering the combination of the three evaluation strategies. We can observe that the ratio improves when the combination of the different strategies is considered: for nested-loop join the mean of this ratio is 0.10, whereas for the combination of strategies the mean is 0.07; this is because the optimizer searches in a larger space of possibilities, increasing the chance of finding better query plans.
 \end{itemize}

In general, we may state that the results show a significant improvement in the evaluation cost for the optimized queries with respect to the worst-case and median-case query orderings. This property holds for synthetic and real-world ontologies. However, for synthetic ontologies we notice that for star-shaped queries, the difference between the median cost and the optimal cost is very small; this indicates that the form of the query may influence the  cost improvement achieved by the optimizer.

\begin{figure}[h]
\centering
\includegraphics[width=0.6\linewidth]{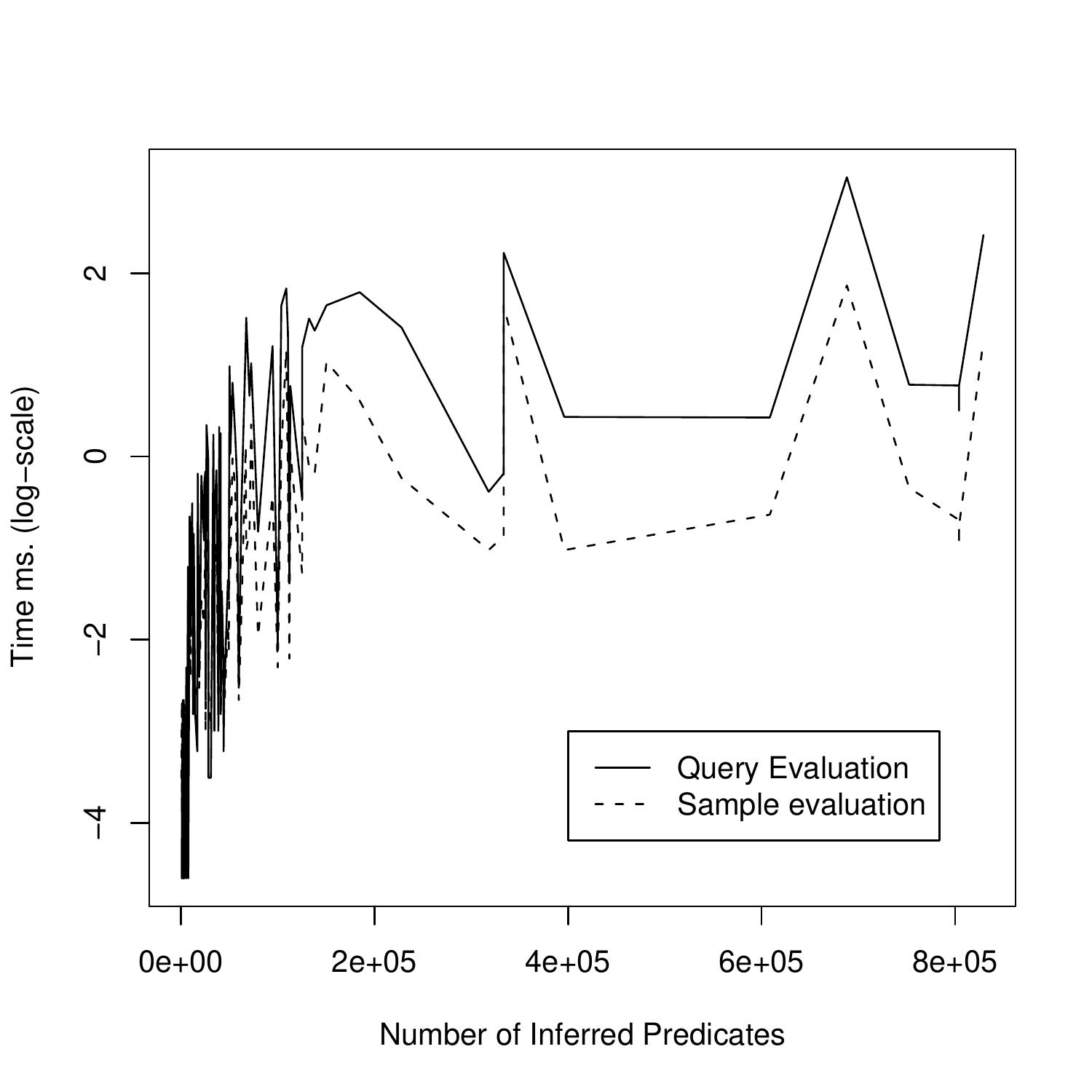}
\caption{Sampling Conjunctions - Query Eval. time and Sample Eval. time vs. \# Inf. Pred.}
\label{Figure3}
\end{figure}

Finally, we would like to point out that we also studied the use of an adaptive sampling technique for the cost estimation of the conjunction of two or more predicates (instead of System R cost model). Although,  the sampling technique gives a similar correlation result than the combination of sampling and System R cost model, the time required to compute the cost estimation may be as large as the  time needed to evaluate the query. In Figure \ref{Figure3}, we can observe that the time difference is marginal. 

\section{Conclusions and Future Work}
We have developed a cost model that combines System R and adaptive sampling techniques. Adaptive sampling is used to estimate data that do not exist a priori, data related to the cardinality and cost of intensional rules in the DOB. The experimental results show that our proposed techniques produce in general a significant improvement in the evaluation cost for the optimized query. 


Currently, we are developing a hybrid optimization mechanism that combines Magic Sets and our cost-based technique; the idea is to first identify a good ordering, and then apply Magic Sets rewritings to reduce the program that evaluates the query. Initial experiments show that this combined solution outperforms the behavior of each individual technique.
 
We plan to apply similar optimization techniques for conjunctive queries to DL ontologies. Initially, we will work on ABox queries extending the the techniques proposed in \cite{sirin06}. In a next stage, we will consider mixed TBox and ABox conjunctive queries.
 \bibliography{bibliotlp}
\bibliographystyle{acmtrans}
\end{document}